\documentclass[seceq]{ptptex}

\usepackage{epsf}

\title{%        %You can use \\ for explicit line-break
The origin of the soft excess in high $L/L_{Edd}$ AGN
}
\author{%       %Use \sc for the family name
Chris {\sc Done}$^{1,}$\footnote{email: chris.done@durham.ac.uk}
}

\inst{%         %Affiliation, neglected when [addenda] or [errata]
$^1$Department of Physics, University of
  Durham, Durham DH1 3LE, UK
\\
}

\recdate{%      %Editorial Office will fill in this.
}

\abst{% %this abstract is neglected when [addenda] or [errata] 

We discuss the origin of the soft X--ray excess seen
in AGN. There are clear advantages to models where this arises from
atomic processes in partially ionised material rather than where it is
a true continuum component. However, current data cannot distinguish
between models where this material is seen in reflection or
absorption, even for the archetypal 'reflection dominated' AGN
MCG--6--30--15. Instead,
we give physical arguments on the ionisation structure
of X--ray illuminated material which exclude a reflection origin if
the disc is in hydrostatic equilibrium. The same physical processes
strongly favour an absorption origin for the soft excess, giving a
more messy picture of the accretion environment. This implies that
these apparently 'reflection dominated' AGN are not good places to
test GR, but they do give insight into the spectra expected from the
first QSO's in the early Universe.

}

\begin{document}

\maketitle

\section{Introduction}

Many high mass accretion rate AGN show X--ray spectra which rise
smoothly below 1~keV above the extrapolated 2--10~keV emission
\cite{rf:1}, equivalent to a fixed temperature of $\sim
0.1-0.2$~keV. This is far too high a temperature to be simply the high
energy tail of the accretion disc emission, and its lack of relation
to the underlying disc temperature argues strongly against it being
Compton scattered disc emission\cite{rf:2}. The apparently
fixed temperature is much easier to explain if it arises from atomic
rather than continuum processes. One potential physical association is
with the large increase in opacity between 0.7--3~keV due to
OVII/OVIII and Fe L shell absorption.  However, the soft excess is
observed to be fairly featureless, so if it is atomic in origin then
there must be a strong velocity shear in order to Doppler smear the
characteristic atomic features into a pseudo--continuum.

Partially ionised material with a strong velocity shear can produce
the soft excess in two different geometries, one where the material is
optically thick and out of the line of sight, seen via reflection
(e.g. from an accretion disc). Alternatively, the material can be
optically thin and in the line of sight, seen in absorption (e.g. a
wind above the disc). We discuss each of these possiblities in detail
below, but show that the model degeneracies mean that {\em both} can
fit the data even of the archetypal 'reflection dominated' AGN
MCG--6--30--15.

Instead we use physical arguments on the 'fine tuning' of the
ionisation parameter to show that the absorption model is strongly
favoured, and that reflection from a hydrostatic disc cannot produce
the soft X--ray excess\cite{rf:3}

\section{Reflection}

The increase in absorption opacity between 0.7--3~keV means a decrease
in reflection between these energies, or equivalently, a rise in the
reflected emission below 0.7~keV, producing a soft excess. This
continuum reflection is enhanced by emission lines from the partially
ionised material, especially OVII/VIII Ly $\alpha$ at 0.6--0.7~keV as
well as Ly$\alpha$ lines from C,N and Fe L transitions\cite{rf:4}. 
Such optically thick reflection from material in the
inner disc is strongly smeared by the disc velocity field and such
models can match the shape of the soft excess. Importantly the
parameters required for the relativistic smearing of the soft excess
can be the {\em same} as those required to produce the associated iron
K$\alpha$ line emission, though objects with the highest
signal--to--noise require {\em multiple} reflection components to fit
the spectra\cite{rf:5}\tocite{rf:8}.

\begin{figure}[b]
\epsfxsize = 0.95\textwidth 
\centerline{\epsfbox{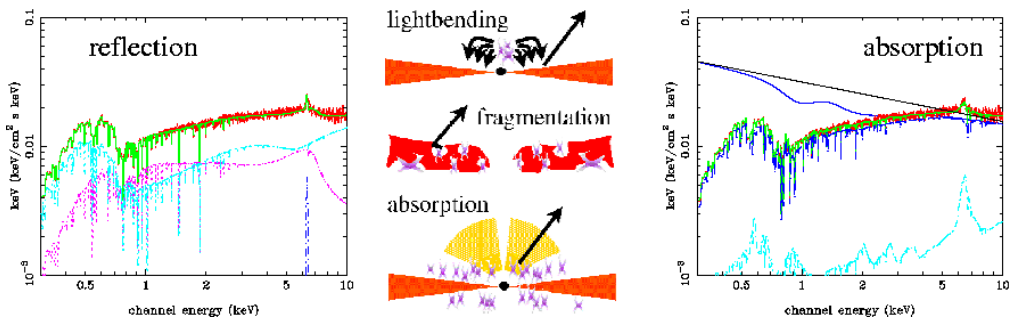}}
\caption{The spectra in the left and right hand panels show 
the deconvolved XMM-Newton data from MCG--6--30--15 using the
reflection and absorption models, respectively. The middle panel shows
sketch geometries for a reflection dominated system (lightbending and disc
fragmentation) compared to absorption in an outflow from the disc.}
\label{fig:1}
\end{figure}

However, the parameters derived from these models can be uncomfortably
extreme, with the amount of smearing implying extreme Kerr spacetime,
with perhaps also extraction of the spin energy of the black hole
\cite{rf:8,rf:9}. Also, the size of the soft
X-ray excess can be much larger than expected for a reflection origin
with isotropic illumination\cite{rf:10}. The
objects with these large soft X--ray excesses then require anisotropic
illumination models, e.g. where the X--ray source is extremely close
to the black hole so that lightbending suppresses the observed direct
continuum flux and enhances the disc illumination\cite{rf:7}.
Alternatively, the disc might fragment into inhomogeneous
regions which hide a direct view of the intrinsic source flux\cite{rf:5,rf:6}.
These reflection geometries are sketched in the
middle panel of Fig 1.

MCG--6--30--15 is the archetypal object which shows all these
features. The left panel of Fig 1 shows the XMM data for this object,
deconvolved with an intrinsic power law, its reflection from two
different ionisation and velocity smeared reflectors both with twice
solar iron abundance, an additional narrow neutral iron line, and
three narrow warm absorber systems to account for the complex
absorption seen at low energies\cite{rf:11}. This 
fits the data ($\chi^2=2327.2/1874$) but requires that the intrinsic
power law (with $\Gamma=2.17$) is not visible, and that one of the
reflectors has extreme smearing ($r_{in}=1.28$, $r_{out}=3.5$, with
highly centrally concentrated emissivity)\cite{rf:9}.

\section{Absorption}

The same physical process of the opacity increase between 0.7--3~keV
can also produce the soft excess via absorption, plausibly from a wind
from the accretion disc\cite{rf:2}, as sketched in Fig
1. Again, relativistic velocity shear is required to smear out the
characteristic atomic features into a pseudo--continuum but the
difference between here is that these motions
are no longer Keplarian, so cannot be used to simply infer the inner
disc radius (and hence black hole spin).

The right hand panel of Fig 1 shows this model fit to the XMM-Newton
data. The model description is the same as for the reflection fits,
except that one of the smeared reflectors is replaced with the smeared
absorption model, and there is no need for an additional narrow iron
line. This gives a similarly good fit to the data
($\chi^2=2215.6/1877$) but now the intrinsic power law is seen
($\Gamma=2.31$), and the remaining reflector has $\Omega/2\pi=0.5$ and
is not extremely smeared ($R_{in}=25$, with emissivity consistent with
the expected gravitational energy release i.e. $\propto r^{-3}$).

Plainly the 0.3--10~keV spectral fits alone cannot distinguish between
a reflection and absorption origin, not even when variability is
included\cite{rf:7,rf:12}. Nor can
data at higher energies as the two models also make very similar
predictions for the 10--30~keV flux of $\sim 3\times 10^{-11}$ ergs
cm$^{-2}$ s$^{-1}$ from the observed 2--10~keV flux of $4.5\times
10^{-11}$ ergs cm$^{-2}$ s$^{-1}$. This is not the case in other
objects, such as 1H0707-496 and especially PG~1211+104, 
where Suzaku HXD data may break the model
degeneracies\cite{rf:10}. 
The difference here is the additional complexity due to
the narrow warm absorber systems at low energies which gives more
freedom in deconvolving the underlying spectrum.

\section{Reflection from a disc}

We can try instead to break the model degeneracies using physical
plausibility arguments as opposed to observational data. Both
absorption and reflection require the same basic ionization conditions
i.e. partially ionised Oxygen to produce the big jump in opacity at
0.7~keV (equivalently $\xi\sim 10^3$ where $\xi=L/nr^2$ is the
photoionisation parameter).  This
may arise rather naturally in an absorption geometry if the material
is in some sort of pressure balance\cite{rf:13}. 
The front of the cloud is heated
by the X--ray illumination, so expands, so its density is low and
ionisation is high. Further into the cloud the heating is less intense
so the material is cooler, so must be denser to be in constant
pressure. The lower ionisation finally allows ion species to exist,
dramatically enhancing the cooling and hence increasing the density.
This rapid transition means the cloud has to contract, which may mean
this neutral material is strongly clumped. A line of sight though the
cloud includes only the highly ionized front edge (invisible) and the
partially ionized transition region, which has an average value of
$\log\xi\sim 3$ across a region with column of $\sim 10^{22-23}$
cm$^{-2}$\cite{rf:13}.

However, the {\em same} rapid change in ionisation state in an X--ray
illuminated, hydrostatic disc produces a similarly stratified vertical
structure but has very different observational consequences.  Again
the rapid transition from completely ionised to mostly neutral occurs
over a column of $10^{22-23}$ cm$^{-2}$, i.e. an optical depth of $\le
0.01$. Thus the zone with the 'correct' ionisation parameter to
produce the soft excess is only a very small fraction of the total
disc photosphere ($\tau=0\to 1$), so the soft excess is very much
smaller than that produced from constant density reflection models
which can arbitrarily set $\xi=10^3$ over the entire photosphere (see
Fig 2).  Since even the constant density models require a reflection
dominated geometry to match the strongest soft excesses seen, this
means that reflection from a hydrostatic disc simply cannot be the
origin of these soft excesses\cite{rf:3}.

\begin{figure}[t]
\begin{tabular}{cc}
\epsfxsize = 0.45\textwidth \epsfbox{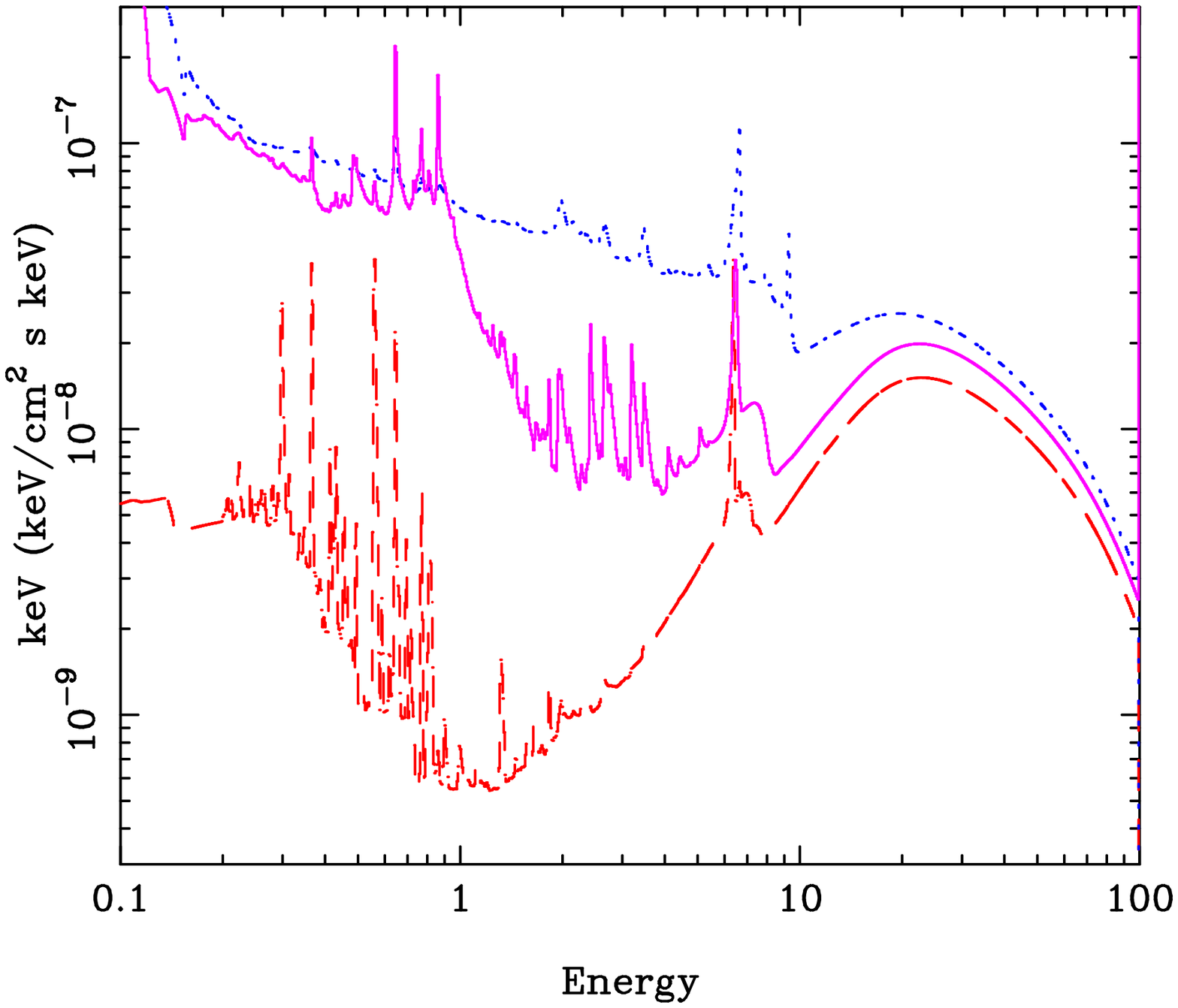} &
\epsfxsize = 0.45\textwidth \epsfbox{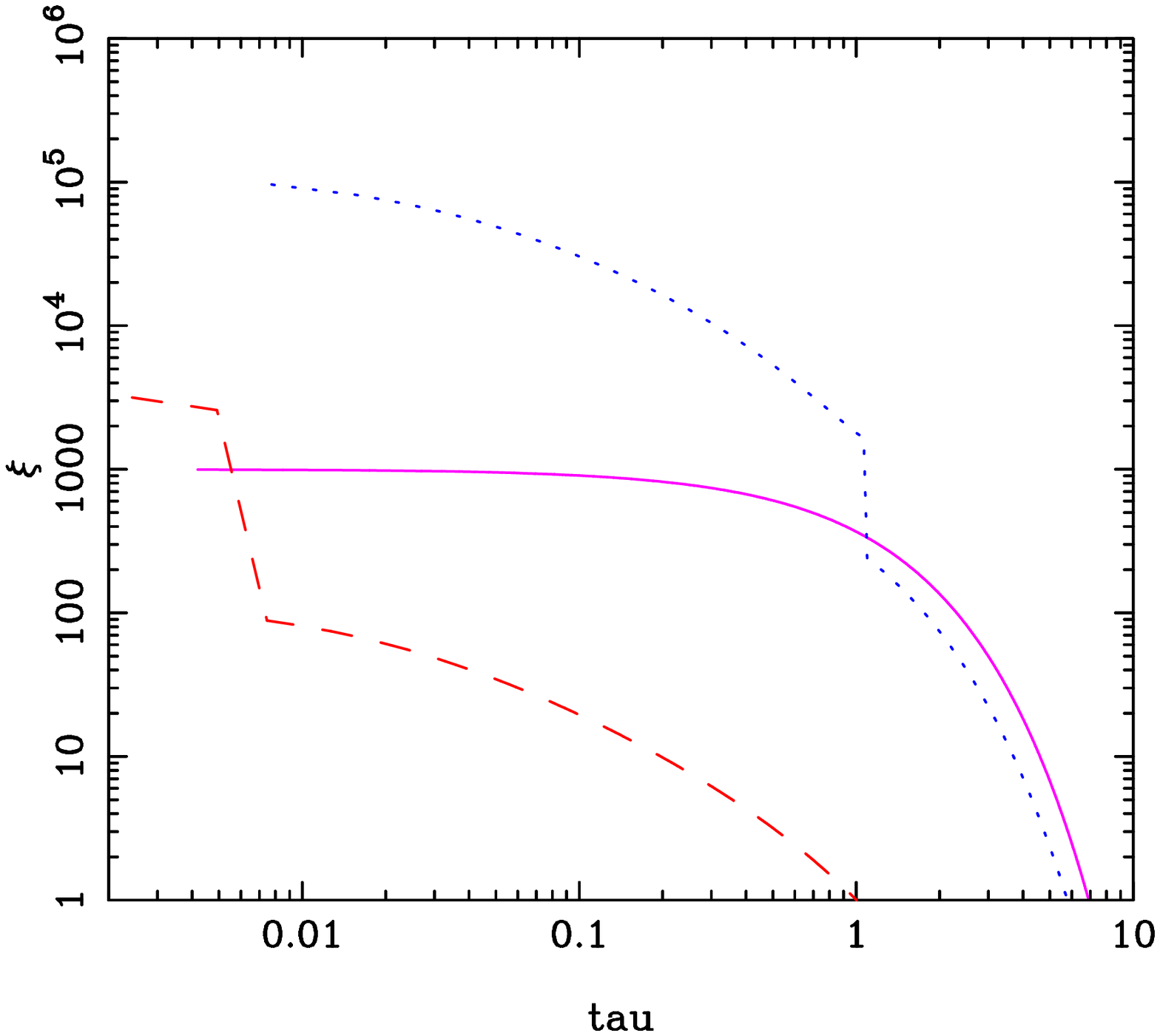}
\end{tabular}
\caption{The left hand panel shows the reflected spectra produced from
a disc in hydrostatic equilibrium at two different mass accretion
rates (dashed red, dotted blue) compared to that from a constant
density slab (solid magenta). The right hand panel shows the corresponding ionisation
parameter.  }
\label{fig:1}
\end{figure}

\section{Conclusions}

The rapid transition from complete ionisation to mostly neutral 
material is characteristic of any pressure balance condition. This
means that the partially ionised zone required for both atomic models
of the soft X--ray excess is limited to a column of $10^{22-23}$
cm$^{-2}$. This strongly supports the 
optically thin, absorption geometry as the origin 
for the soft X--ray excess, and this more messy picture of these high
mass accretion rate AGN means they are probably not good places to
test GR.


\begin{thebibliography}{99}
%%%%%%%%%%%%%%%%%%%%%%%%%%%%%%%%%%%%%%%%%%%%%%%%%%%%%%%%%%%%%
% Some macros are available for the bibliography:
%   o for general use
%      \JL : general journals          \andvol : Vol (Year) Page
%   o for individual journal 
%      \PR  : Phys. Rev.               \PRL : Phys. Rev. Lett.
%      \NP  : Nucl. Phys.              \PL  : Phys. Lett.
%      \JMP : J. Math. Phys.           \CMP : Commun. Math. Phys.
%      \PTP : Prog. Theor. Phys.       \JPSJ: J. Phys. Soc. Jpn.
%      \JP  : J. of Phys.              \NC  : Nouvo Cim.
%      \IJMP: Int. J. Mod. Phys.       \ANN : Ann. of Phys.
% Usage:
%   \PR{D45,1990,345}            ==> Phys.~Rev.\ {\bf D45} (1990), 345
%   \JL{Phys.~Lett.,A30,1981,56} ==> Phys.~Lett.\ {\bf A30} (1981), 56
%   \andvol{B123,1995,1020}      ==> {\bf B123} (1995), 1020
%%%%%%%%%%%%%%%%%%%%%%%%%%%%%%%%%%%%%%%%%%%%%%%%%%%%%%%%%%%%%


\bibitem{rf:1} {Porquet} D., {Reeves} J.~N., {O'Brien} P., {Brinkmann} W.,
  2004, A\& A, 422, 85

\bibitem{rf:2} Gierli{\'n}ski M., Done C., 2004, MNRAS, 349, L7 

\bibitem{rf:3} Done C., Nayakshin S., 2001, MNRAS, 328, 616 

\bibitem{rf:4} Ross R.~R., {Fabian} A.~C., 2005, MNRAS, 358, 211

\bibitem{rf:5} Fabian A.~C., et al., 2002, MNRAS, 331, L35

\bibitem{rf:6} {Fabian} A.~C., {Miniutti} G., {Iwasawa} K., {Ross} R.~R., 2005, MNRAS, 361,
  795

\bibitem{rf:7} {Miniutti} G., {Fabian} A.~C., 2004, MNRAS, 349, 1435

\bibitem{rf:8} {Crummy} J., {Fabian} A.~C., {Gallo} L., {Ross} R.~R., 2006, MNRAS, 365, 1067

\bibitem{rf:9} Wilms J., et al., 2001, MNRAS, 328, L27 

\bibitem{rf:10} Sobolewska M.~A., Done C., 2007, MNRAS, 374, 150 

\bibitem{rf:11}  Turner, A.~K., Fabian, 
A.~C., Lee, J.~C., \& Vaughan, S.\ 2004, MNRAS, 353, 319 

\bibitem{rf:12} {Gierli{\'n}ski} M., {Done} C., 2006, MNRAS, L64

\bibitem{rf:13} {Chevallier} L., et al., 2006, A\& A, 449, 493


\end{thebibliography}
\end{document}